\documentclass[aps,pre,onecolumn,superscriptaddress,letterpaper]{revtex4}
\usepackage{mathrsfs}
\usepackage{amsmath}
\usepackage{bm}
\usepackage{graphicx}
\usepackage{hyperref}

\begin{document}           


\title{Half Spectral, an Another General Method for Linear Plasma Simulation}
\author{Hua-sheng XIE\footnote{Email: huashengxie@gmail.com}}
\affiliation{Institute for Fusion Theory and Simulation, Zhejiang
University, Hangzhou, 310027, PRC}
\date{\today}

\begin{abstract}
There are two usual computational methods for linear (waves and
instabilities) problem: eigenvalue (dispersion relation) solver and
initial value solver. In fact, we can introduce an idea of the
combination of them, i.e., we keep time derivative dt term (and
other term if have, e.g., kinetic dv term), but transform the linear
spatial derivatives dx term to ik, which then can reduce the
computational dimensions. For example, most (fluid and kinetic)
normal mode problems can be reduced from treating cumbersome PDEs to
treating simple ODEs. Examples for MHD waves, cold plasma waves and
kinetic Landau damping are given, which show to be extremely simple
or even may be the simplest method for simulating them. [I don't
know whether this idea is new, but it seems very interesting and
useful. So, I choose making it public.]
\end{abstract}



\maketitle

\section{Introduction and Basic Idea}\label{sec:intro}
It's well known in plasma physics community that for linear (waves
and instabilities) problems we have two usual computational methods:
eigenvalue (dispersion relation) solver and initial value solver.
For the former, we transform the linear time derivative $\partial
/\partial t$ and spatial derivatives $\nabla$ to spectral space
using $-i\omega$ and $i{\bf k}$; for the latter, we solve the
original equations directly. In some simple case, the eigenvalue
method can be reduced to analytic tractable form, e.g., many well
known dispersion relations are this type. However, numerical
solutions are always OK (except some singularity cases).

Usually, the eigenvalue method is not intuitive and one needs be
good at theoretical derivations; the conventional initial value
method is complicated in computation and cumbersome in data
analysis. Typically, the eigenvalue method can give all solutions of
the system, while the initial value method can only give the
$\gamma_{max}$ (most unstable) solution.

Can we combine these two methods? The answer is yes. We can keep
time derivative $\partial /\partial t$ term (and other term if have,
e.g., kinetic $\partial /\partial {\bf v}$ term), but transform the
linear spatial derivatives $\partial /\partial {\bf x}$ term to
$i{\bf k}$, which is still an initial value method, but is solved in
half (only spatial not temporal) spectral space, then also has
characteristics of eigenvalue method. For example, we can highlight
the non-$\gamma_{max}$ solutions.

We will show how to do it with examples for normal mode problems in
Sec.\ref{sec:normal}. Since this method has been used for linear
inhomogeneous eigenvalue and nonlinear problem by previous
researchers, we will just give some necessary descriptions with
citations in Sec.\ref{sec:eigen}.

\section{Normal Mode Paradigms}\label{sec:normal}
In most literatures, normal mode and eigenmode are treated as a same
concept since they are very similar. But, here, we distinguish them
\begin{description}
  \item[Normal mode:] homogeneous\cite{dw}, without boundary conditions;
  \item[Eigen mode:] inhomogeneous or with boundary conditions,
  all possible solutions of the system can be expressed by
  proper sum of eigenmodes.
\end{description}

Using Half Spectral (HS) method, many normal mode problems can be
reduced from PDEs to ODEs. Then can be solved extreme easily. While,
unfortunately, for eigen mode problems, the equations are in
lower-dimensions but still PDEs.

\subsection{Simple example}
We construct a simple example to show how to use this method. The
original equations
\begin{equation} \label{eq:simple_hs_1}
    \left\{ \begin{aligned}
  \frac{\partial {f_1}}{\partial t} + {u_a}\frac{\partial {f_1}}{\partial x} + {u_b}\frac{\partial {f_2}}{\partial x} &=
  0, \\
  \frac{\partial {f_2}}{\partial t} + {u_b}\frac{\partial {f_1}}{\partial x} + {u_a}\frac{\partial {f_2}}{\partial x} &=
  0.
    \end{aligned} \right.
\end{equation}
Half spectral solves
\begin{equation} \label{eq:simple_hs_hs}
    \left\{ \begin{aligned}
  \partial {f_1}/\partial t &=  - (ik{u_a}{f_1} + ik{u_b}{f_2}), \\
  \partial {f_2}/\partial t &=  - (ik{u_a}{f_1} + ik{u_b}{f_2}.
    \end{aligned} \right.
\end{equation}
The dispersion relation
\begin{equation} \label{eq:simple_hs_dr}
    \left\{ \begin{aligned}
  (\omega  - k{u_a}){f_1} = k{u_b}{f_2}, \\
  (\omega  - k{u_a}){f_2} = k{u_b}{f_1}.
    \end{aligned} \right.
\end{equation}
gives, ${\omega _ \pm } = k({u_a} \pm {u_b}),~{f_1} =  \pm {f_2}$.

If we assume the initial values ${f_1} = y{f_2}$, the ratio of
$\omega_\pm$ is $x$ and $1-x$ respectively, then $x-(1-x)=y
\Rightarrow x = (y + 1)/2$, then the ratio of the amplitudes for
$\omega_\pm$ is
\begin{equation} \label{eq:simple_hs_Amp}
{{{{\rm{A}}_ + }} \over {{{\rm{A}}_ - }}} = \left| {{x \over {1 -
x}}} \right| = \left| {{{y + 1} \over {y - 1}}} \right|,
\end{equation}
which means we can control the amplitudes of each modes of the
system exactly by set the proper initial values.

\begin{figure}
  \includegraphics[width=14cm]{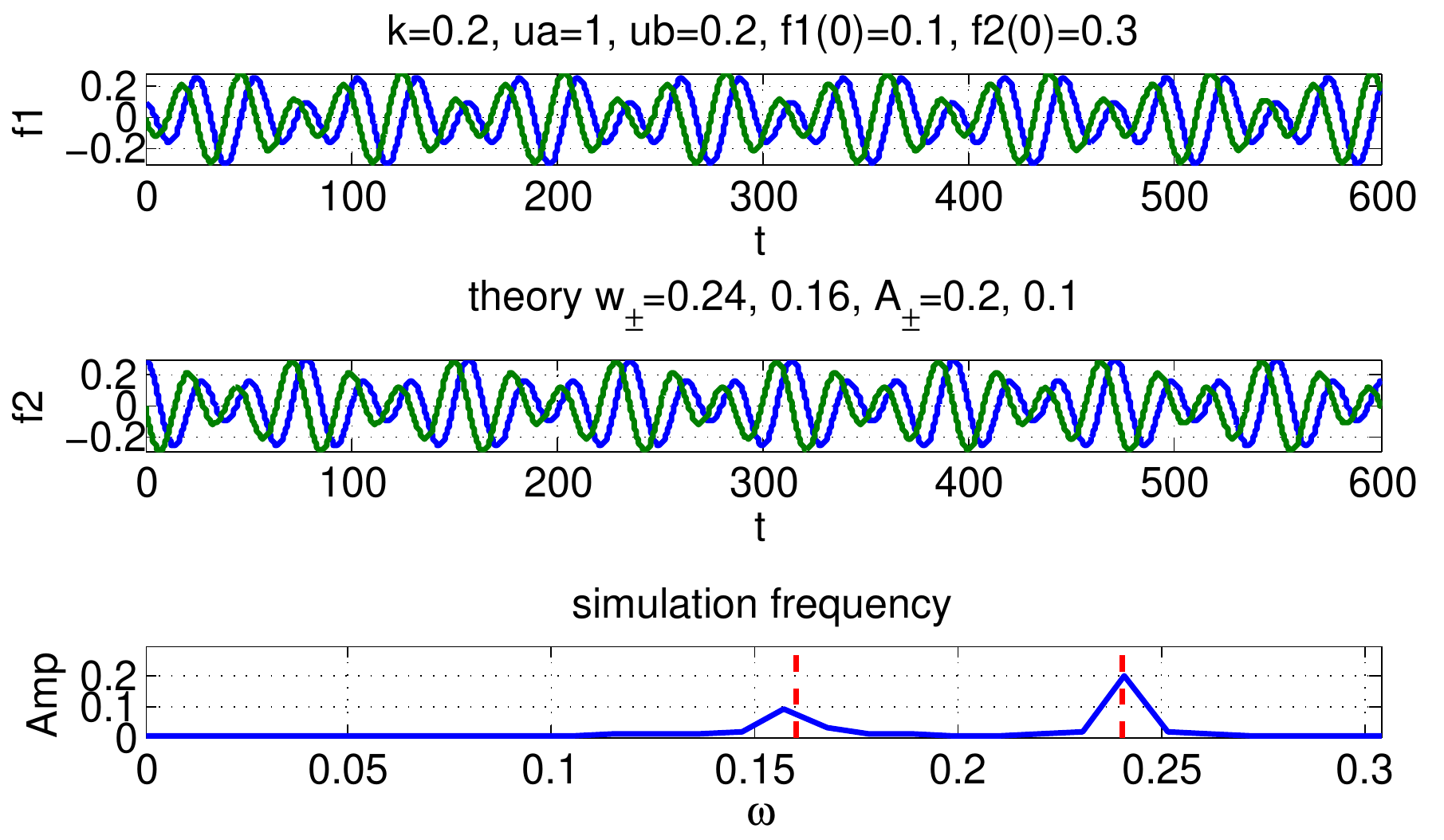}\\
  \caption{RK4 to solve eq.(\ref{eq:simple_hs_hs}), as example for half spectral method, red line is dispersion relation solutions}\label{fig:simple_hs}
\end{figure}

A 4th order Runge-Kutta simulation of eq.(\ref{eq:simple_hs_hs}) is
shown in Fig.\ref{fig:simple_hs}. We can see the frequencies and
amplitudes of each modes ($\omega_\pm=0.24,0.16$, $A_\pm=0.2,0.1$)
are exact as predict. A small mismatch should be caused by the
numerical discrete.

\subsection{ES1D kinetic problem}
Usually, we have two initial value method to simulate kinetic
problem (e.g., Landau damping), i.e., Vlasov continuity solver and
PIC method.

To show that half spectral method is not only for fluid problem, we
give a kinetic example. At this subsection, the electrostatic 1D
Landau damping and bump-on-tail simulations are given.

The linearized equations (ion immobile) are
\begin{equation} \label{eq:es1d}
    \left\{ \begin{aligned}
  \frac{\partial \delta f}{\partial t} &=  - ikv\delta f + \frac{e}{m}E\frac{\partial {f_0}}{\partial v}, \\
  ik\delta E &=  - e\int {\delta fdv}.
    \end{aligned} \right.
\end{equation}
which gives
\begin{equation} \label{eq:es1d_df}
\delta f = \frac{ieE}{m}\frac{{\partial _v}{f_0}}{\omega - kv},
\end{equation}
contains both normal mode (which is independent with initial value)
and ballistic mode (which is brought by initial value) or phase
mixing (see e.g., \cite{Krall1973}).

\begin{figure}
  \includegraphics[width=12cm]{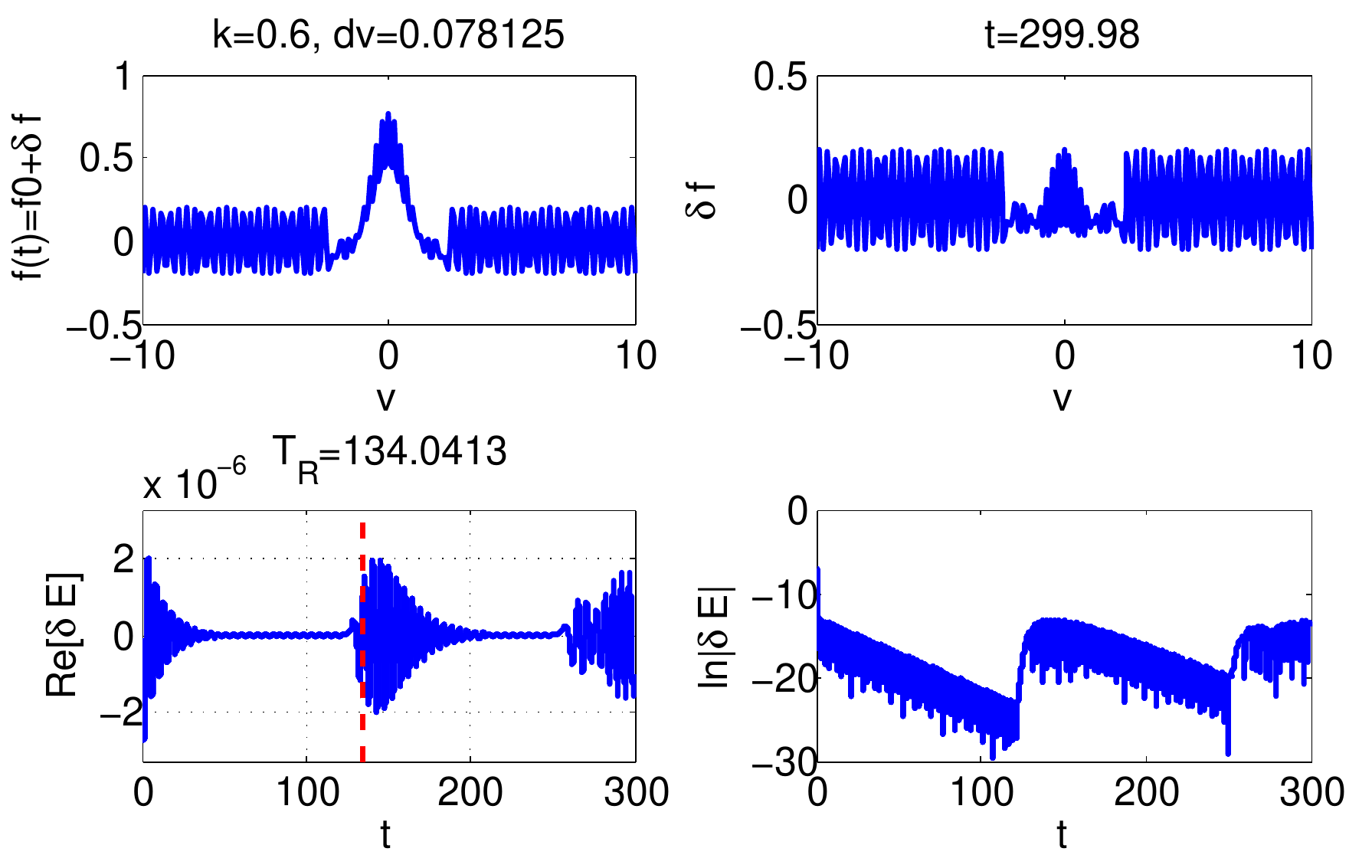}\\
  \caption{Solve eq.(\ref{eq:es1d}) for Landau damping}\label{fig:landaudamping}
\end{figure}

\begin{figure}
  \includegraphics[width=10cm]{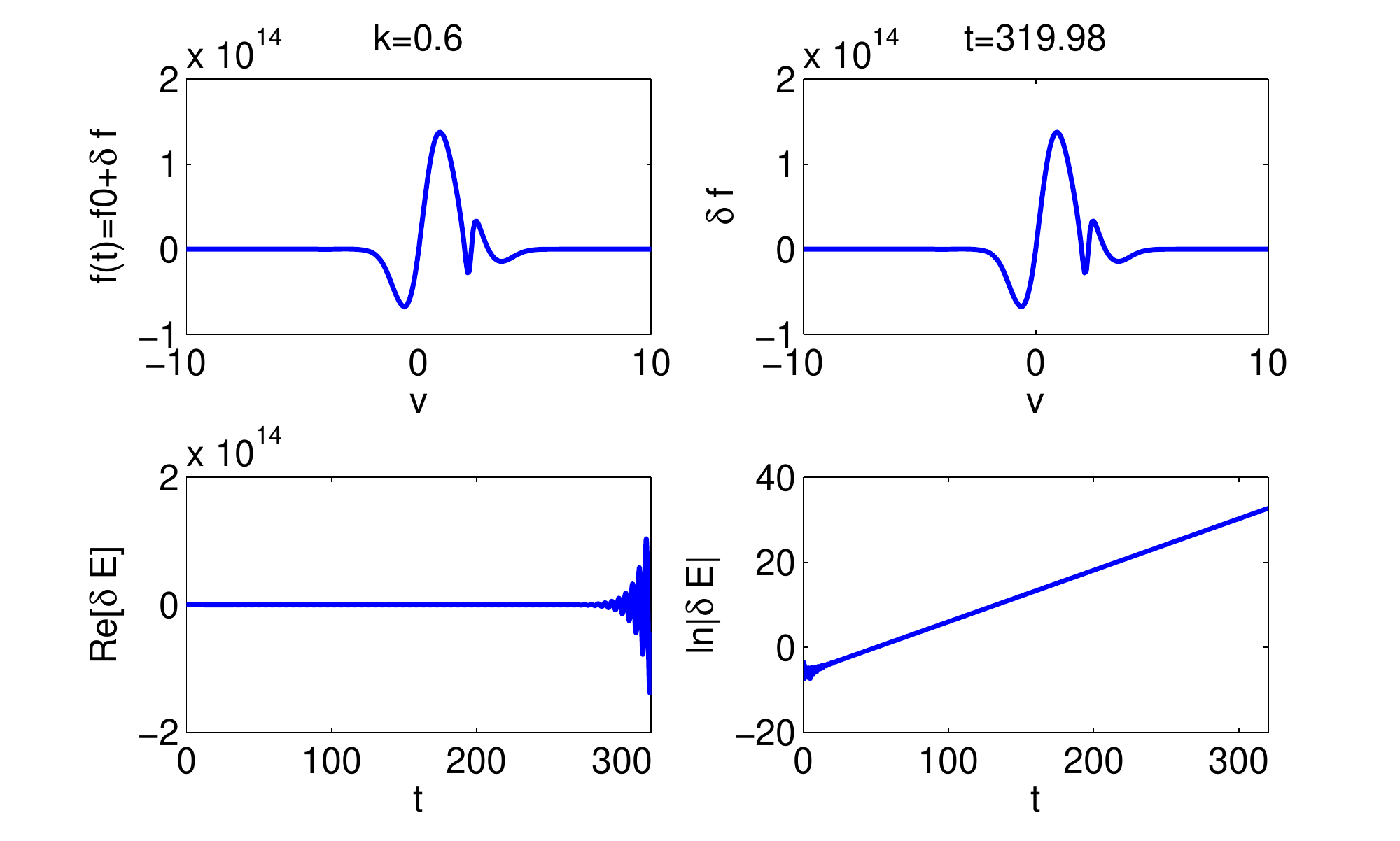}\\
  \caption{Solve eq.(\ref{eq:es1d}) for beam-plasma instability}\label{fig:bumpontail}
\end{figure}

The initial distribution function $f_0$ can be any form, e.g.,
Maxwellian gives Landau damping, bump-on-tail gives beam-plasma
instabilities.

Fig.\ref{fig:landaudamping} shows the simulation of Landau damping.
One can find very similar results (especially the fourth panel) from
Vlasov continuity simulation. However, we should notice an
unfavorite recurrence effect\cite{Cheng1976} caused by discrete
$\Delta v$, which is also found in half spectral simulation. The
recurrence time $T_R=2\pi/k\Delta v$.

Fig.\ref{fig:bumpontail} is the simulation of bump-on-tail problem.
Since we treat linear problem, the unit of the amplitude can be
arbitrary large.

One can find, comparing with continuity solver and PIC method, using
half spectral method for Landau damping simulation is extreme simple
and can be more accurate. Comparing with numerical or analytical
dispersion relation solver, half spectral method does not need treat
troublesome or confusing integral contours. Especially, when the
initial distribution function is not standard and the dispersion
relation is hard to solve, the half spectral simulation can gives an
reasonable solution for benchmark more complicated codes.

\subsection{MHD waves}
For MHD waves, we solve
\begin{equation} \label{eq:mhd}
    \left\{ \begin{aligned}
  {{\partial \delta \rho } \over {\partial t}} &=  - i{\rho _0}{\bf{k}} \cdot \delta {\bf{u}}, \\
  {\rho _0}{{\partial \delta {\bf{u}}} \over {\partial t}} &= {i \over {{\mu _0}}}({\bf{k}} \times \delta {\bf{B}}) \times {{\bf{B}}_0} - i{\bf{k}}v_s^2\delta \rho, \\
  {{\partial \delta {\bf{B}}} \over {\partial t}} &= i{\bf{k}} \times (\delta {\bf{u}} \times {{\bf{B}}_0}).
    \end{aligned} \right.
\end{equation}
where ${{\bf{B}}_0} = (0,0,{B_0})$, ${\bf{k}} = (k\sin \theta
,0,k\cos \theta )$, $v_s^2 = \gamma {p_0}/{\rho _0} = \gamma
k{T_0}/m$, $v_A^2 = B_0^2/{\mu _0}{\rho _0}$, ${v_p} = \omega /k$.

Three solutions are fast mode, slow mode and shear Alfv\'en wave
(one can find introductions of them in textbooks, e.g.,
\cite{Gurnett2005})
\begin{equation} \label{eq:mhd_waves}
    \left\{ \begin{aligned}
  v_p^2 &= {1 \over 2}(v_A^2 + v_s^2) + {1 \over 2}{\left[ {{{(v_A^2 - v_s^2)}^2} + 4v_A^2v_s^2{{\sin }^2}\theta } \right]^{1/2}}, \\
  v_p^2 &= {1 \over 2}(v_A^2 + v_s^2) - {1 \over 2}{\left[ {{{(v_A^2 - v_s^2)}^2} + 4v_A^2v_s^2{{\sin }^2}\theta } \right]^{1/2}}, \\
  v_p^2 &= v_A^2{\cos ^2}\theta.
    \end{aligned} \right.
\end{equation}

\begin{figure}
  \includegraphics[width=15cm]{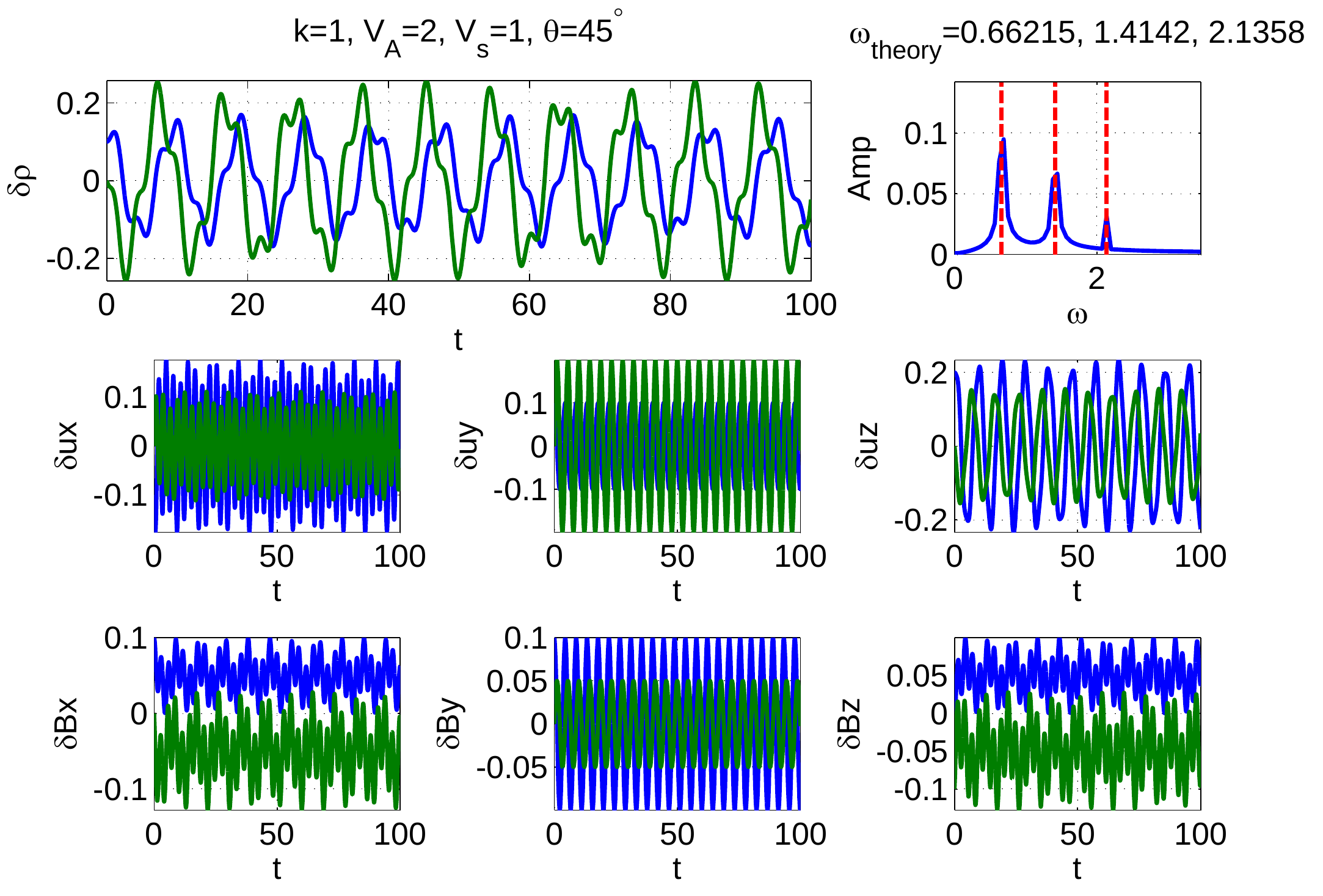}\\
  \caption{Solve eq.(\ref{eq:mhd}) for ideal MHD waves, red line is dispersion relation solutions}\label{fig:mhd_waves}
\end{figure}

A simulation result is shown in Fig.\ref{fig:mhd_waves}. Again, we
can find the simulation result exactly matches the theoretical
solutions. The simulation is intuitive. The frequency signal in the
second panel is taken from $\delta\rho$ and $\delta u_y$. If we only
use $\delta\rho$ signal, the shear Alfv\'en wave solution will
vanish.

If one want to go non-ideal MHD, e.g., including resistivity or
using anisotropic pressure $\delta p_{\parallel} \neq \delta
p_{\bot}$, but wouldn't like to do analytical derivations, then half
spectral method is an useful choice: it is very simple, intuitive
and can give solutions exact enough.

Since all information for linear perturbation variables is kept in
the simulation, we can use them for many more deeply analysis, e.g.,
the polarization and so on.

\subsection{EM cold plasma waves}
Equations are
\begin{equation} \label{eq:cold_waves}
    \left\{ \begin{aligned}
  \frac{\partial \delta {\bf{v}_s}}{\partial t} &= \frac{e_s}{m_s}[\delta {\bf{E}}+\delta {\bf{v}_s}\times\delta {\bf{B}}], \\
  \frac{\partial \delta {\bf{E}}}{\partial t} &= ic^2{\bf{k}} \times \delta {\bf{B}} - \delta{\bf{J}}/\epsilon_0, \\
  \frac{\partial \delta {\bf{B}}}{\partial t} &= -i{\bf{k}} \times \delta {\bf{E}}.
    \end{aligned} \right.
\end{equation}
where $\delta{\bf{J}}=\sum_{s} n_{s0}e_s\delta {\bf{v}_s}$. And,
${{\bf{B}}_0} = (0,0,{B_0})$, ${\bf{k}} = (k\sin \theta ,0,k\cos
\theta )$, $\omega_{cs}=e_sB_0/m_s$ and
$\omega_{ps}=n_sq_s^2/\epsilon_0m_s$.

\begin{figure}
  \includegraphics[width=15cm]{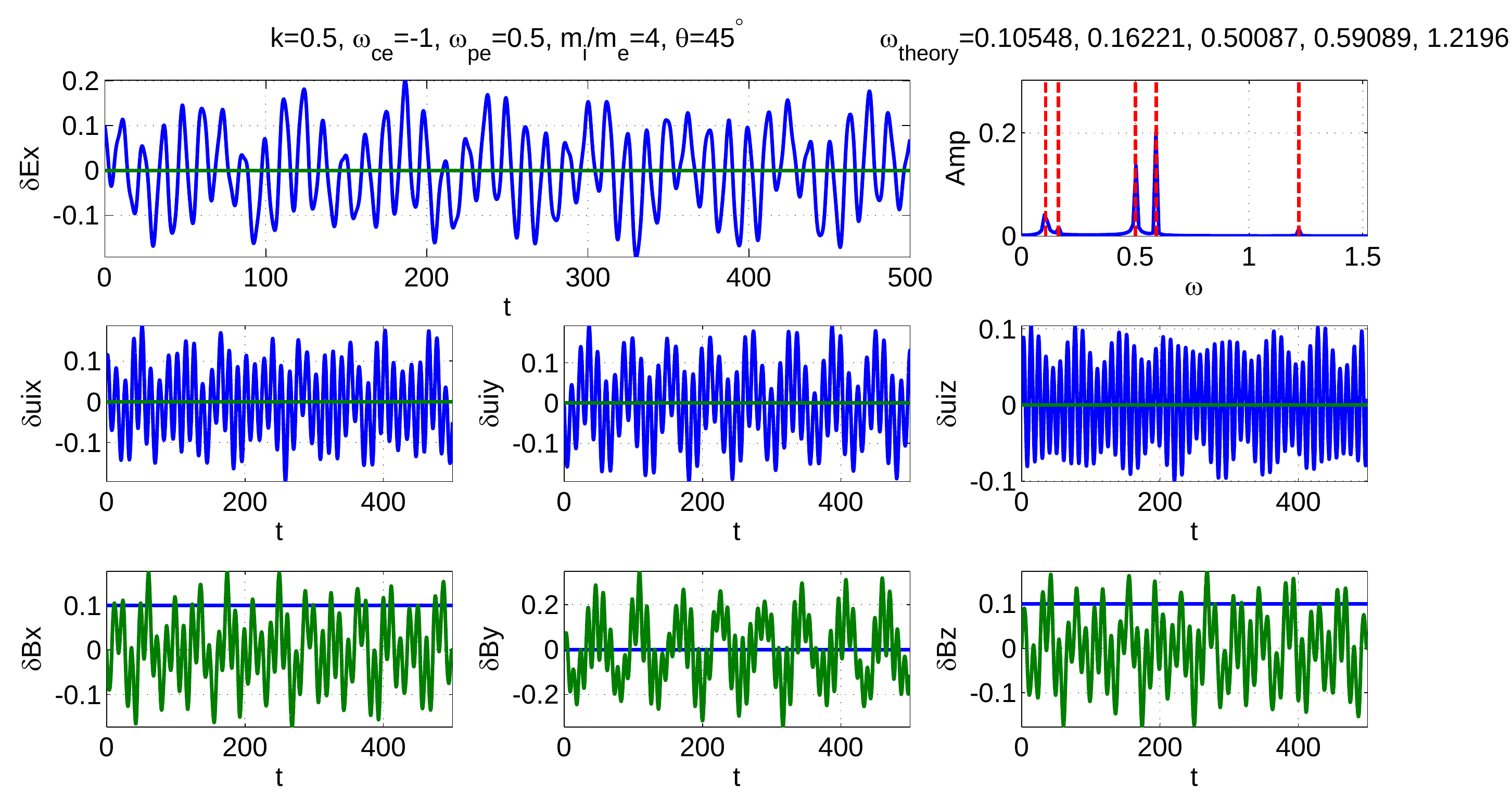}\\
  \caption{Solve eq.(\ref{eq:cold_waves}) for EM cold plasma waves, red line is dispersion relation solutions}\label{fig:em_cold_waves}
\end{figure}

For one ion species, the final dispersion relation can be reduced
(with heavy calculations) to a fifth order equation for $\omega^2$
(see \cite{Swanson2003} for details). While, using half spectral
simulation, this is very easy. A result is shown in
Fig.\ref{fig:em_cold_waves}

If we have more than one ion species or with beams, the dispersion
relation can be very headache even though the problem seems not that
complicated. While, using half spectral method, the problem is
indeed still very easy.

\subsection{Summary}
For normal mode problem, half spectral method may be the simplest
method for simulating them, which is intuitive, simple and also
exact enough. For analytical difficult problems, this method can not
only be an intuitive tool but also has practical usages.

\section{Eigenmode and Nonlinear Problems}\label{sec:eigen}
As claimed at the title, half spectral method can be a general
(linear) plasma simulation method. So, we also need discuss the
eigenmode problem and some nonlinear treatments. While, it is found
in the literatures that previous researchers have given many
examples of this. So, here we just give a short description and
mention some citations.

\subsection{Tearing mode}
Eigenmode problems are similar. We take collisional tearing mode as
example here.

A simulation matches the half spectral idea is given by Lee and
Fu\cite{Lee1986} (see also citations of that paper). For simulation,
we solve
\begin{equation} \label{eq:tearing}
    \left\{ \begin{aligned}
    {{\partial \delta \rho } \over {\partial t}} &=  - {{\partial \rho } \over {\partial x}}\delta {u_x} - \rho {{\partial \delta {u_x}} \over {\partial x}} - i\alpha \rho \delta
  {u_z}, \\
  {{\partial \delta {u_x}} \over {\partial t}} &=  - {\beta  \over {2\rho }}{{\partial \delta p} \over {\partial x}} - {{{B_z}} \over \rho }{{\partial \delta {B_z}} \over {\partial x}} - {1 \over \rho }{{\partial {B_z}} \over {\partial x}}\delta {B_z} + i\alpha {{{B_z}} \over \rho }\delta
  {B_x}, \\
  {{\partial \delta {u_z}} \over {\partial t}} &=  - i\alpha {{\beta \delta p} \over {2\rho }} + {1 \over \rho }{{\partial {B_z}} \over {\partial x}}\delta
  {B_x}, \\
  {{\partial \delta {B_x}} \over {\partial t}} &= i\alpha {B_z}\delta {u_x} + {1 \over {{R_m}}}{{{\partial ^2}\delta {B_x}} \over {\partial {x^2}}} - {{{\alpha ^2}} \over {{R_m}}}\delta
  {B_x}, \\
  {{\partial \delta {B_z}} \over {\partial t}} &=  - {{\partial {B_z}} \over {\partial x}}\delta {u_x} - {B_z}{{\partial \delta {u_x}} \over {\partial x}} + {1 \over {{R_m}}}{{{\partial ^2}\delta {B_z}} \over {\partial {x^2}}} - {{{\alpha ^2}} \over {{R_m}}}\delta
  {B_z}, \\
  {{\partial \delta p} \over {\partial t}} &=  - {{\partial p} \over {\partial x}}\delta {u_x} - \gamma p{{\partial \delta {u_x}} \over {\partial x}} - i\alpha \gamma p\delta {u_z}.
    \end{aligned} \right.
\end{equation}
where parameter are $\alpha  = kl$, $\beta  = 2{\mu _0}{p_\infty
}/B_\infty ^2$, ${R_m} = {v_A}l/\eta$, $\gamma$. The normalization
unit are ${B_\infty }$, ${\rho _\infty }$, ${v_A} = B_\infty ^2/{\mu
_0}{\rho _\infty }$, ${p_\infty }$, $l$, ${t_0} = l/{v_A}$.

The results are very well in that paper\cite{Fu1995}. So, we can
trust that half spectral method is also good for eigenmode problem.

A bad thing is that, for eigenmode problem, we still need solve
PDEs.

\subsection{Hasegawa-Mima equation}
In fact, an usual way for nonlinear drift wave turbulence by
Hasegawa-Mima equation\cite{Hasegawa1978} simulation is doing in
spectral space, which is exact the half spectral idea of this
manuscript. A very detailed introduction can be found in Waltz's
lecture notes\cite{Waltz1986}. To give a rough impression, one of
the equations is shown below
\begin{equation}\label{eq:hm}
    \frac{d}{dt}\phi_k(t)=(-i\omega_{0k}+\gamma_k)\phi_k(t)+\frac{1}{2}\sum_{k_1k_2}\delta(k-k_1-k_2)V_{kk_1k_2}\phi_{k_1}(t)\phi_{k_2}(t).
\end{equation}

We should comment here, for nonlinear problem, we need sum all $k$
modes and the complex conjugate should also be kept.

\section{Summary and Comments}
In this manuscript, we discussed the idea of half spectral method,
and showed that it can be a general simulation method. However, this
idea maybe not new, especially for eigenmode and nonlinear problem,
many previous researchers used it. I don't know whether this method
is new for normal mode problems yet. And the name {\it half
spectral} using here is just for convenient.

For tokamak (or other problems with strong guide field) simulation,
a similar method is called flux tube (e.g., \cite{Peeters2009}),
which using the idea of symmetry to reduce dimensions because the
physics is mainly along magnetic field line then we can take it as
one coordinate. Poloidal and toroidal mode numbers $m$ and $n$ are
often used in flux tube simulation. I haven't checked whether the
equations for flux tube simulation are the same for half spectral
simulation yet. They may be exact the same or at least very similar.

But, we can tell that spectral or pseudo spectral method are
different from this half spectral method, because that they need
transform back to real space. For a same simulation, spectral or
pseudo spectral method can be alternative by other discrete methods
(e.g., finite difference) and won't influent the simulation results.
While, half spectral method is independent on discrete methods.

This idea is partly inspired by ``{\it A 2D Hasegawa-Mima Model of
Electrostatic Drift Wave Turbulence}" simulation by Deng ZHAO (PKU,
2011) and P239C course project (UCI, 2010) ``{\it Cold plasma-warm
beam interaction}" given by Prof. Liu CHEN. The basic idea is the
same. The new here is that we find this method can be generalized.

If we do not care that whether this idea is new or not, we can find
half spectral method is simple, interesting and useful. MATLAB codes
for this manuscript are given in attached files.


\begin{thebibliography}{99}
\bibitem{dw} For drift wave, if the gradient parameter $\partial ln(f)/\partial
x$ is constant, we also treat it as normal mode.
\end{thebibliography}
\end{document}